\documentclass[aps,prb,twocolumn,showpacs,amsmath,amssymb,superscriptaddress]{revtex4}
\usepackage{graphicx}
\usepackage{amssymb}
\usepackage{natbib}
\usepackage{color}

\usepackage{epstopdf}

\newcommand{\beq}{\begin{eqnarray}}
\newcommand{\eeq}{\end{eqnarray}}

\begin{document}
	
\title{Local breaking of four-fold rotational symmetry by short-range magnetic order in heavily overdoped Ba(Fe$_{1-x}$Cu$_{x}$)$_{2}$As$_{2}$}

\author{Weiyi Wang}
\affiliation{Department of Physics and Astronomy, Rice University, Houston, Texas 77005, USA}
\author{Yu Song}
\email{Yu.Song@rice.edu}
\affiliation{Department of Physics and Astronomy, Rice University, Houston, Texas 77005, USA}
\author{Ding Hu}
\affiliation{Department of Physics and Astronomy, Rice University, Houston, Texas 77005, USA}
\affiliation{Center for Advanced Quantum Studies and Department of Physics, Beijing Normal University, Beijing 100875, China}
\author{Yu Li}
\affiliation{Department of Physics and Astronomy, Rice University, Houston, Texas 77005, USA}
\author{Rui Zhang}
\affiliation{Department of Physics and Astronomy, Rice University, Houston, Texas 77005, USA}
\author{L. W. Harriger}
\affiliation{NIST Center for Neutron Research, National Institute of Standards and Technology, Gaithersburg, Maryland 20899, USA}
\author{Wei Tian}
\affiliation{Quantum Condensed Matter Division, Oak Ridge National Laboratory, Oak Ridge, Tennessee 37831, USA}
\author{Huibo Cao}
\affiliation{Quantum Condensed Matter Division, Oak Ridge National Laboratory, Oak Ridge, Tennessee 37831, USA}
\author{Pengcheng Dai}
\email{pdai@rice.edu}
\affiliation{Department of Physics and Astronomy, Rice University, Houston, Texas 77005, USA}
\affiliation{Center for Advanced Quantum Studies and Department of Physics, Beijing Normal University, Beijing 100875, China}
	
\begin{abstract}
We investigate Cu-doped Ba(Fe$_{1-x}$Cu$_x$)$_2$As$_2$ with transport, magnetic susceptibility, and elastic neutron scattering measurements. In the heavily Cu-doped regime where long-range stripe-type antiferromagnetic order in BaFe$_2$As$_2$ is suppressed, Ba(Fe$_{1-x}$Cu$_x$)$_2$As$_2$ (0.145 $\leq x \leq$ 0.553) samples exhibit spin-glass-like behavior in magnetic susceptibility and insulating-like temperature dependence in electrical transport. Using elastic neutron scattering, we find stripe-type short-range magnetic order in the spin-glass region identified by susceptibility measurements. 
 The persistence of short-range magnetic order over a large doping range in Ba(Fe$_{1-x}$Cu$_x$)$_2$As$_2$ likely arises from local arrangements of Fe and Cu that favor magnetic order, with Cu acting as vacancies relieving magnetic frustration and degeneracy. These results indicate locally broken four-fold rotational symmetry, suggesting that stripe-type magnetism is ubiquitous in iron pnictides.
\end{abstract}
	
	
\maketitle

%\section{INTRODUCTION}

The parent compounds of iron pnictides such as BaFe$_2$As$_2$ and NaFeAs exhibit stripe-type antiferromagnetic (AF) order below $T_{\rm N}$ that breaks both spin-rotational symmetry and four-fold rotational symmetry of the underlying crystalline lattice \cite{PDai}. However, a tetragonal-to-orthorhombic structural transition occurs at $T_{\rm s}$ with $T_{\rm N}\leq T_{\rm s}$, preemptively breaking four-fold rotational symmetry of the crystal and result in an Ising-nematic state in the region $T_{\rm N}\leq T\leq T_{\rm s}$ \cite{RMFernandes12_PRB}. The Ising-nematic state and associated fluctuations have been implicated in superconducting pairing of iron pnictides \cite{HHKuo16_Science}, although nature of the nematic state is still under debate \cite{RMFernandes_NP}.

Superconductivity can be induced by substituting Fe with transition metals such as Co and Ni, which also suppresses the magnetic and structural phase transitions \cite{GRStewart,PDai,PCCanfield10}. In the overdoped regime where both the magnetic and structural transitions are suppressed, the system maintains average four-fold rotational symmetry without long-range magnetic order, although inelastic neutron scattering revealed substantial stripe-type fluctuations even in non-superconducting overdoped BaFe$_{1.7}$Ni$_{0.3}$As$_2$ \cite{MWang13_NC}. For BaFe$_{2-x}TM_x$As$_2$ ($TM =$ Co, Ni, Cu), while Co- and Ni-doping result in superconducting domes with optimal $T_{\rm c}\sim20$ K, optimal $T_{\rm c}\sim 2$ K or no superconductivity is observed in $A$Fe$_{2-x}$Cu$_x$As$_2$ ($A$ = Ba, Sr) \cite{NNi10,MGKim,YJYan}. This contrast points to the inadequacy of a simple rigid band picture \cite{HWadati10,TBerlijn12,PVilmercati16} and highlights differences between dopants \cite{RKraus13,MGKim12}.     

Compared to $A$Fe$_{2-x}$Cu$_x$As$_2$, superconductivity with optimal $T_{\rm c}=11.5$ K is observed in NaFe$_{1-x}$Cu$_{x}$As \cite{AFWang,GTan16}. With increasing Cu concentration ($x\gtrsim10\%$), insulating-like transport and short-range magnetic order develop, evolving towards an insulator with long-range magnetic order and Fe-Cu ordering near $x\approx50\%$ \cite{YSong}. Evolution from metallic to insulating/semiconducting transport is also observed in Ba(Fe$_{1-x}$Cu$_{x}$)$_2$As$_2$ ($x\geq$ 0.145), Sr(Fe$_{1-x}$Cu$_{x}$)$_2$As$_2$ ($x\geq$ 0.06) \cite{YJYan}, Fe$_{1.01-x}$Cu$_x$Se ($x\geq$ 0.03) \cite{THuang,AJWilliams}, and Fe$_{1+\delta-x}$Cu$_x$Te ($x\geq$ 0.06) \cite{HWang,PNValdivia}. The insulating transport in NaFe$_{1-x}$Cu$_{x}$As is a result of electron correlations \cite{YSong,CYe} facilitated by ordering of Fe and Cu into quasi-1D chains \cite{YSong}. In contrast, disorder is suggested to be responsible for the insulating transport in Fe$_{1-x}$Cu$_x$Se \cite{SChadov}.   

In NaFe$_{1-x}$Cu$_x$As with short-range magnetic order, temperature-dependence of the magnetic order parameter is broad indicative of spin-glass (SG) behavior, commonly observed in doped strongly-correlated materials \cite{JAMydosh15}. SG behavior seen in magnetization measurements is also reported for other heavily Cu-doped iron pnictides \cite{AFWang,YJYan} and chalcogenides \cite{THuang,AJWilliams,HWang,PNValdivia}, pointing to the possible presence of short-range magnetic order. Importantly, in both $A$(Fe$_{1-x}$Cu$_{x}$)$_{2}$As$_2$ and NaFe$_{1-x}$Cu$_x$As where SG behavior is observed, doped Cu are in nonmagnetic 3$d^{10}$ configuration \cite{YJYan,YSong} and therefore any SG or short-range magnetic order must be due to Fe. While the long-range magnetic and Fe-Cu orders in the ideal NaFe$_{0.5}$Cu$_{0.5}$As compound lacks four-fold rotational symmetry, there is so far no evidence of magnetic order or four-fold symmetry breaking in other iron pnictides and chalcogenides in the non-superconducting heavily overdoped regime.      

In this Rapid Communication, we investigate heavily overdoped Ba(Fe$_{1-x}$Cu$_{x}$)$_{2}$As$_2$ with electrical transport, magnetic susceptibility, and elastic neutron scattering measurements. Similar to other Cu-doped iron pnictides and chalcogenides, Ba(Fe$_{1-x}$Cu$_{x}$)$_{2}$As$_{2}$ shows insulating and SG behaviors in the heavily overdoped regime. Using elastic neutron scattering, we discovered the presence of stripe-type short-range AF order over a large region of the Ba(Fe$_{1-x}$Cu$_{x}$)$_{2}$As$_2$ phase diagram ($10\%\lesssim x\lesssim50\%$). The spin-spin correlation length along the in-plane longitudinal direction is found to be much  longer than along the in-plane transverse direction, revealing locally broken four-fold rotation symmetry. Our discovery of short-range magnetic order in Ba(Fe$_{1-x}$Cu$_{x}$)$_{2}$As$_2$ reveals an inherent instability towards stripe-type magnetic order and highlights the role of magnetic frustration in iron pnictides.

%\section{EXPERIMENTAL DETAILS}
Single crystal of Ba(Fe$_{1-x}$Cu$_x$)$_2$As$_2$ samples were prepared using the same self-flux method as for BaFe$_{2-x}$Ni$_x$As \cite{YChen}. The Cu substitution levels reported in this paper were determined by inductively coupled plasma atomic-emission spectroscopy (ICP). Samples with nominal Cu concentrations of 10, 20, 30, 50 and 70\% were prepared, resulting in actual Cu concentrations of $x$ = 14.5, 25.4, 31.6, 44.7 and 55.3\%. $x\approx$ 50\% is the highest doping level that is achievable with our synthesis method.

Magnetic susceptibility measurements were carried out using a commercial superconducting quantum interference device magnetometer from Quantum Design, measurements were taken on warming with applied field $\mu_{0}H = 1$T perpendicular to the $c$-axis. In-plane electrical resistivity measurements were carried out using the standard four-probe method on a commercial physical property measurement system from Quantum Design. 

Elastic neutron scattering experiments were carried out using the SPINS triple-axis spectrometer (TAS) ($x = 0.145, 0.254, 0.316, 0.447, 0.553$) at the NIST Center for Neutron Research, and the HB-1A TAS ($x = 0.553$) at the High Flux Isotope Reactor, Oak Ridge National Laboratory. We used pyrolytic graphite [PG(002)] monochromators and analyzers in all experiments. At SPINS, the monochromator is vertically focused and the analyzer is flat with fixed $E_{\rm f}$ = 3.7 meV. At HB-1A, the monochromator is vertically focused with fixed incident neutron energy $E_{\rm i}$ = 14.6 meV and the analyzer is flat. Be filter and PG filter were used at SPINS and HB-1A to avoid contamination from higher-order neutrons. The collimations of guide-40$^\prime$-sample-40$^\prime$-open and 40$^\prime$-40$^\prime$-sample-40$^\prime$-80$^\prime$ were used at SPINS and HB-1A, respectively. Single-crystal neutron diffraction experiment on a $x$ = 0.316 sample was carried out using the four-circle diffractometer HB-3A at the High Flux Isotope Reactor, Oak Ridge National Laboratory. The data was measured at 5 K with a neutron wavelength of 1.005 $\text{\AA}$ from a bent Si(331) monochromator using an Anger camera detector. 
% confirm with Rui about experiment details.

While Ba(Fe$_{1-x}$Cu$_x$)$_2$As$_2$ is tetragonal for $x\gtrsim$ 0.05 \cite{MGKim}, our results are reported using the orthorhombic structural unit cell of BaFe$_2$As$_2$ ($a\approx b\approx 5.6$ {\AA} and $c\approx12.9$ {\AA}) \cite{QHuang} [Fig. 1(b)]. The momentum transfer $\textbf{Q}=H\textbf{a}^\ast+K\textbf{b}^\ast+L\textbf{c}^\ast$ is denoted as $\textbf{Q}=(H,K,L)$ in reciprocal lattice units (r.l.u.), where $H$, $K$, $L$ are Miller indices and ${\bf a}^\ast = \hat{{\bf a}}2\pi/a$, ${\bf b}^\ast=\hat{{\bf b}}2\pi/b$ and ${\bf c}^\ast=\hat{{\bf c}}2\pi/c$. In this notation, magnetic Bragg peaks in the parent compound BaFe$_2$As$_2$ appears at $\textbf{Q}=(1,0,L)$ with $L=1,3,5,\cdots$. Samples were aligned in the $[H,0,0]\times[0,0,L]$ scattering plane, which allows scans along $H$ and $L$ centered at $\textbf{Q}=(1,0,L)$. To carry out scans of the magnetic peak along $K$ at $\textbf{Q}=(1,0,1)$, the $x=0.553$ sample was also studied in the $[H,0,H]\times[0,K,0]$ scattering plane [Fig. 1(d)].

%\section{RESULTS}

\begin{figure}[t]
	\includegraphics[scale=.4]{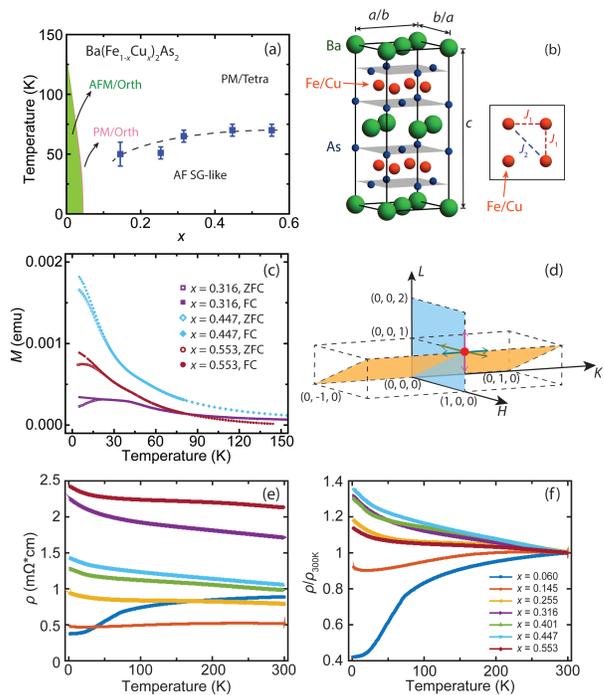}
	\caption{
		(Color online) (a) The phase diagram of Ba(Fe$_{1-x}$Cu$_{x}$)$_2$As$_2$. The region with $x\lesssim$ 0.1 is obtained from ref. \citenum{MGKim}. The onset temperature of short-range magnetic order on the overdoped side are measured with an energy resolution $\Delta E\approx0.1$ meV using SPINS. Error bars are estimated uncertainties of the onset temperature of short-range magnetic order. (b) The crystal structure of Ba(Fe$_{1-x}$Cu$_{x}$)$_2$As$_2$, using the orthorhombic structural unit cell of BaFe$_2$As$_2$. Exchange couplings are defined for the Fe/Cu-plane on the right of the unit cell. (c) Temperature dependence of magnetic susceptibility measured with ZFC and FC. The data are magnetic moment induced by applied magnetic field, with 1 emu = 10$^{-3}$ A$\cdot$m$^2$. (d) Schematic of $[H,0,L]$ and $[H,K,H]$ scattering planes that allow scans centered at $\textbf{Q}=(1,0,1)$ along $H$, $L$ and $K$ directions. (e) Temperature dependence of the in-plane electrical resistivity for Ba(Fe$_{1-x}$Cu$_{x}$)$_2$As$_2$ single crystals. (f) Temperature dependence of the resistivity normalized to its room temperature value.
	}
\end{figure}

Figure 1(a) summarizes the overall phase diagram of Ba(Fe$_{1-x}$Cu$_{x}$)$_2$As$_2$. The stripe-type AF order in BaFe$_2$As$_2$ is suppressed at $x\approx5\%$ \cite{MGKim}. For samples with $x\gtrsim$ 0.1, we detected the presence of short-range magnetic order occurring at the stripe-type ordering vector, with the onset temperatures determined from elastic neutron order parameter measurements on SPINS [Fig. 3(b)-(f)]. Magnetic susceptibility results for Ba(Fe$_{1-x}$Cu$_{x}$)$_2$As$_2$ single crystals ($x = 0.316$, $0.447$ and $0.553$) with zero-field-cooling (ZFC) and field-cooling (FC) are shown in Fig. 1(c). The separation between ZFC and FC susceptibilities occurring at low temperatures indicates SG-like behavior, similar to other heavily Cu-doped iron pnictides and chalcogenides \cite{AFWang,YJYan,HWang,AJWilliams,THuang,PNValdivia}. 

The temperature dependence of the in-plane resistivity for Ba(Fe$_{1-x}$Cu$_{x}$)$_2$As$_2$ single crystals are shown in Figure 1(e), and resistivity normalized to its room temperature value ($\rho/\rho_{300 {\rm K}}$) is shown in Fig. 1(f). The evolution from metallic to insulating-like transport with increasing Cu substitution is similar to Sr(Fe$_{1-x}$Cu$_x$)$_2$As$_2$ \cite{YJYan}. Despite the insulating-like temperature dependence, the largest measured resistivity for all samples is of the order ${\rm m}\Omega\cdot{\rm cm}$, smaller than resistivity in NaFe$_{1-x}$Cu$_x$As with $x\approx50\%$ by three orders of magnitude \cite{YSong}. 

\begin{figure}[t]
	\includegraphics[scale=.4]{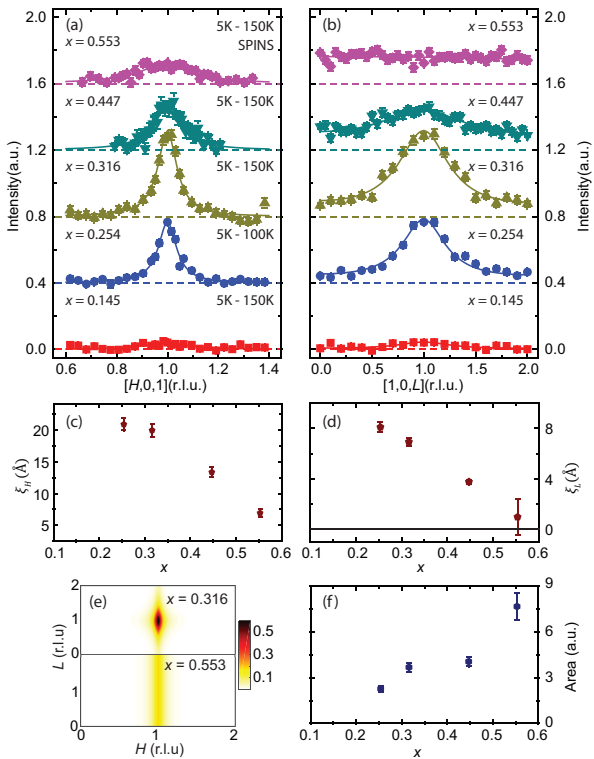}
	\caption{
		(Color online) (a) High-temperature-background-subtracted elastic scans centered at $\textbf{Q}=(1,0,1)$ along the $H$ direction for Ba(Fe$_{1-x}$Cu$_{x}$)$_2$As$_2$ with $x = 0.553$, $0.447$, $0.316$, $0.254$ and $0.145$. Corresponding scans along the $L$ direction are shown in (b). Solid lines are fits with the lattice sum of Lorentzian peaks multiplied by the magnetic form factor. The results for different dopings are normalized by sample mass. (c) and (d) show spin-spin correlation lengths as a function of doping along $H$ and $L$, respectively. (e) Spin-spin correlation functions in the $[H,0,L]$ plane for $x=0.316$ and 0.553 samples, obtained by multiplying fits along $H$ and $L$. (f) Integrated magnetic intensity in the $[H,0,L]$ plane as a function of doping, obtained from fitting results in (a) and (b). The error bars in (a) and (b) represent statistical error (1 s.d.). The error bars in (c), (d) and (f) are from least square fits (1 s.d.).
	}
\end{figure}

Elastic neutron scattering scans along $H$ and $L$ centered at ${\bf Q}=(1,0,1)$ are summarized in Figs. 2(a)-(b) for Ba(Fe$_{1-x}$Cu$_{x}$)$_2$As$_2$ ($x = 0.145$, 0.254, 0.316, 0.447, 0.553). These results are measured in the $[H,0,L]$ scattering plane using identical configurations on SPINS and backgrounds obtained at high temperatures have been subtracted. The relative intensities have been normalized by sample mass to allow for a direct comparison between different dopings. In the $x=0.145$ sample, a weak peak centered at ${\bf Q}=(1,0,1)$ can be readily seen. Increasing Cu concentration to $x=0.254$, magnetic order becomes significantly stronger. Upon further increasing Cu doping, the magnetic peaks become broader along both directions. Notably in the $x=0.553$ sample, short-range magnetic order becomes almost independent of $L$, forming a rod of diffuse scattering in reciprocal space. 

To extract the doping evolution of spin-spin correlation lengths, we fit the data in Figs. 2(a)-(b) using a lattice sum of Lorentzian peaks multiplied by the magnetic form factor,
$$ I = F^{2}(\mathbf{Q})\sum_{x_{\rm c}} \frac{h (\frac{\Gamma}{2})^{2}}{(x-x_{\rm c})^{2} + (\frac{\Gamma}{2})^{2}},$$
where $F(\mathbf{Q})$ is the dimensionless magnetic form factor, $x$ is either $H$ or $L$, $h$ is the Lorentzian peak height, and $\Gamma$ is the full width at half maximum (FWHM) for the Lorentzian peak. The summation is over $x_{\rm c}=\ldots,-5,-3,-1,1,3,5,\ldots$, corresponding to magnetic Bragg peak positions in BaFe$_2$As$_2$ along $H$ and $L$. Magnetic correlation lengths in units of {\AA$^{-1}$} is obtained through $\xi_{H}=\frac{a}{\pi\Gamma_H}$ along $H$ and $\xi_L=\frac{c}{\pi\Gamma_L}$ along $L$ \cite{MCollins}.

The resulting doping dependence of spin-spin correlation lengths along $H$ and $L$ are respectively shown in Fig. 2(c) and Fig. 2(d). Due to weak magnetic intensity, the correlation lengths for the $x=0.145$ sample cannot be reliably obtained. $\xi_H>\xi_L$ for all measured samples, similar to short-range magnetic order in NaFe$_{1-x}$Cu$_x$As \cite{YSong}. Increasing Cu concentration leads to a decrease of correlation lengths along both $H$ and $L$, in stark contrast to NaFe$_{1-x}$Cu$_x$As where correlation lengths increase with increasing Cu-doping \cite{YSong}. This difference is likely due to Fe and Cu order in NaFe$_{1-x}$Cu$_x$As but not in Ba(Fe$_{1-x}$Cu$_x$)$_2$As$_2$, as discussed below.

\begin{figure}[t]
	\includegraphics[scale = 0.4]{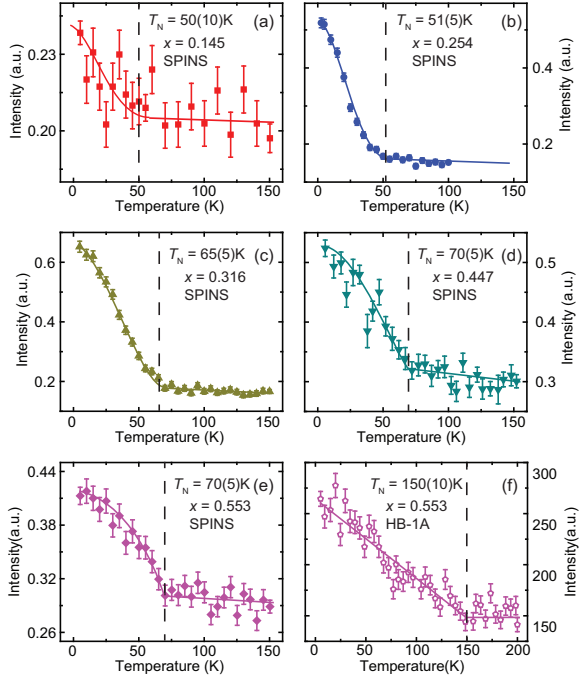}
	\caption{
		(Color online) Temperature dependence of magnetic intensity measured on SPINS at ${\bf Q}=(1,0,1)$ for Ba(Fe$_{1-x}$Cu$_{x}$)$_2$As$_2$ with (a) $x = 0.145$, (b) $x = 0.254$, (c) $x = 0.316$, (d) $x=0.447$ and (e) $x=0.553$. (f) Temperature dependence for the $x=0.553$ sample measured on HB-1A. The error bars represent statistical error (1 s.d.).
	}
\end{figure}

Assuming the spin-spin correlation function in the $[H,0,L]$ plane can be described by multiplying the correlation functions along $[H,0,1]$ and $[1,0,L]$ [obtained from fits in Figs. 2(a)-(b)], the integrated intensity of diffuse scattering in the $[H,0,L]$ plane can be obtained. 2D plots of spin-spin correlation functions in the $[H,0,L]$ plane obtained this way are shown in Fig. 2(e) for $x=0.316$ and $x=0.553$, and doping dependence of the integrated intensity in the $[H,0,L]$ plane is shown in Fig. 2(f). The integrated magnetic signal increases with Cu-doping, suggesting enhanced magnetic moment on Fe due to hole-doping of Cu similar to NaFe$_{1-x}$Cu$_x$As \cite{YSong}. This conclusion also holds when $\xi_K$ is taken into consideration to obtain the integrated volume of diffuse scattering, assuming $\xi_K$ either evolves in a similar fashion as $\xi_H$ and $\xi_L$ with doping or depends weakly on doping. The increase of integrated diffuse scattering is likely a result of the hole-doping effect of Cu, which has 3$d^{10}$ configuration in the heavily-doped regime of iron pnictides \cite{YJYan,YSong,Anand,supplemental}. 

Temperature dependence of magnetic intensity at ${\bf Q}=(1,0,1)$ for different dopings is summarized in Fig. 3, with Fig. 3(a)-(e) obtained on SPINS with an energy resolution $\Delta E \approx 0.1$ meV and Fig. 3(f) measured on HB-1A with $\Delta E \approx 1$ meV. Clear but broad onset of magnetic intensities is observed in all cases, the broad onset is consistent with SG-like behavior revealed by susceptibility measurements in Fig. 2(c). One feature of glassy magnetism is the measured onset temperature of magnetic intensity depends on energy resolution \cite{APMurani,XLu,KBinder}. To see if this is the case in Ba(Fe$_{1-x}$Cu$_x$)$_2$As$_2$, we compare measured temperature dependence on the same $x=0.553$ sample using different instrument energy resolutions, as shown in Fig. 3(e) and (f). With the coarser energy resolution on HB-1A , an onset temperature of $T_{\rm N}\approx150$ K is obtained [Fig. 3(f)] compared to $T_{\rm N}\approx70$ K obtained with finer resolution [Fig. 3(f)], confirming the glassy nature of magnetism. While we only studied the $x=0.553$ sample with different energy resolutions, we anticipate such resolution-dependent onset temperature should be observed for all samples with $0.145\leq x\leq0.553$.  The effect of energy resolution on the onset temperature in Ba(Fe$_{1-x}$Cu$_x$)$_2$As$_2$ is much more significant than what is seen in the cluster spin glass phase of BaFe$_{2-x}$Ni$_x$As$_2$ \cite{XLu}, but comparable to Cu-Mn spin glass alloys \cite{APMurani}. 
   
\begin{figure}[t]
	\includegraphics[scale = 0.4]{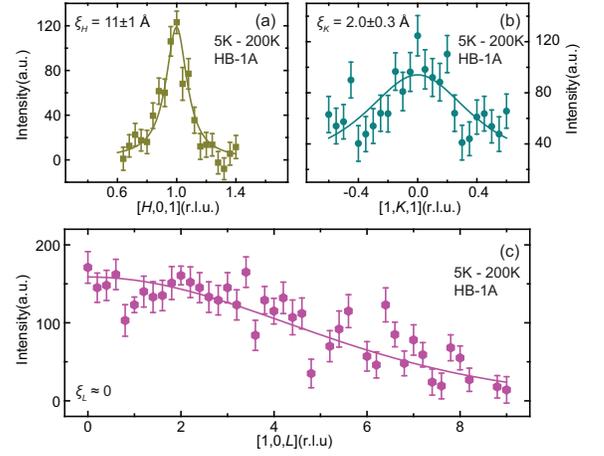}
	\caption{High-temperature-background-subtracted elastic scans along (a) $H$, (b) $K$ and (c) $L$ directions for the $x=0.553$ sample measured on HB-1A. Solid lines are fits with the lattice sum of Lorentzian peaks multiplied by the magnetic form factor. The error bars represent statistical error (1 s.d.).
	}
\end{figure}

While the finding of short-range stripe-type AF order suggests breaking of four-fold rotational symmetry, it is not conclusive evidence. For example, the double-${\bf Q}$ magnetic order that retains four-fold rotational symmetry in hole-doped iron pnicitdes also results in magnetic peaks at the same ordering vector \cite{Avci,Wasser,Allred,Allred2}. A unique way to determine if there is breaking of rotational symmetry for short-range order is to examine the directional dependence of the spin-spin correlation lengths, as done for charge order in cuprates \cite{RComin15_Science}. 

To compare the in-plane spin-spin correlation lengths,  we carried out scans along $H$, $K$ and $L$ directions centered at ${\bf Q}=(1,0,1)$ for the $x=0.553$ sample. The measurements are done by mounting the sample in the $[H,0,L]$ and $[H,K,H]$ scattering planes [Fig. 1(d)], and are carried out using HB-1A. The results are summarized in Fig. 4 and fit using a lattice sum of Lorentzian peaks described above with backgrounds measured at $T=200$ K subtracted from the data. For the in-plane longitudinal direction $H$ [Fig. 4(a)], magnetic signal is broad but relatively well-defined, resulting in a correlation length of $\xi_H\approx11$ {\AA}, in agreement with similar measurement on SPINS [Fig. 2(a)]. For the in-plane transverse direction $K$ [Fig. 4(b)], the signal is considerably broader resulting in $\xi_K\approx2$ {\AA}. We note this large difference of in-plane correlation lengths is intrinsic and unlikely to arise from strain. There is almost no modulation of intensity along $L$ as shown in Fig. 4(c), with the intensity gradually decreasing for increasing momentum transfer following the magnetic form factor, suggesting the magnetic moments are aligned in-plane perpendicular to the ordering vector, similar to heavily-doped NaFe$_{1-x}$Cu$_x$As \cite{YSong}. 

%\section{DISCUSSION}

The highly anisotropic in-plane correlation lengths $\xi_H$ and $\xi_K$ demonstrate that short-range magnetic order in Ba(Fe$_{1-x}$Cu$_x$)$_2$As$_2$ breaks four-fold rotational symmetry of the lattice. The order of correlation lengths in the $x=0.553$ sample is $\xi_H>\xi_K>\xi_L$, identical to short-range magnetic order in NaFe$_{0.61}$Cu$_{0.39}$As \cite{supplemental}, suggesting a similar origin of magnetic order in both systems. Magnetic order in NaFe$_{1-x}$Cu$_x$As becomes long-range when $x\approx50\%$ with Fe and Cu ordering into quasi-1D chains, resulting in super-lattice structural peaks persisting at room temperature \cite{YSong}. In Ba(Fe$_{1-x}$Cu$_x$)$_2$As$_2$, we did not detect any super-lattice peaks for the $x=0.316$ sample on HB-3A, suggesting Cu to be much more disordered. Fe-Cu ordering in NaFe$_{1-x}$Cu$_x$As reduces hopping between Fe ions \cite{YSong}, and its absence likely contributes to the much smaller resistivity in Ba(Fe$_{1-x}$Cu$_x$)$_2$As$_2$.

While there may be weak Fe-Cu ordering in Ba(Fe$_{1-x}$Cu$_x$)$_2$As$_2$ that is hard to pick up in diffraction measurements, it is likely short-range magnetic order can exist even when Cu and Fe are completely disordered. This is because in an disordered system there can be regions with local arrangements of Fe and Cu that are favorable for stabilizing magnetic order. In the $J_1-J_2$ description of magnetism in iron pnictides \cite{QMSi}, the nearest-neighbor coupling $J_1$ and next-nearest-neighbor coupling $J_2$ [Fig. 1(b)] are frustrated, and lead to stripe-type AF order with two degenerate grounds states. Because Cu is non-magnetic and effectively act as vacancies in the heavily-doped regime, for certain arrangements of Fe and Cu [such as, but not limited to, Fe-Cu ordering in NaFe$_{1-x}$Cu$_x$As], the frustration can be relieved and the degeneracy removed, favoring a magnetically ordered state. But because Fe and Cu are overall mostly disordered, such favorable Fe and Cu configurations can only be realized over short length scales, resulting in glassy short-range magnetic order that we observe. The decrease of correlation lengths with increasing Cu concentration [Figs. 2(c) and (d)] is consistent with this picture.

We have observed emergence of stripe-type short-range magnetic order due to the presence of effective vacancies in the system, different from the appearance of stripe-type long-range order in lightly-doped BaFe$_2$As$_2$ through an Ising-nematic state. The robustness of stripe-type magnetism suggests it is an inherent instability of the FeAs-plane and may be the driving force of physics in iron pnictides. Given similar susceptibility and transport behaviors seen in several other Cu-doped iron pnictide and chalcogenide systems \cite{YJYan,HWang,AJWilliams,THuang,PNValdivia}, it is probable stripe-type short-range magnetic order is also present in those systems.       
 
%\section{SUMMARY AND CONCLUSIONS}
In conclusion, we have found short-range magnetic order in Ba(Fe$_{1-x}$Cu$_x$)$_2$As$_2$ (0.145 $\leq x \leq$ 0.553) over an extremely large doping range. Different in-plane correlation lengths of the short-range magnetic order point to locally broken four-fold rotational symmetry. Our finding suggests stripe-type magnetism to be a robust ground state in iron pnictides.

%\section{ACKNOWLEDGMENTS}
The authors gratefully acknowledge Dr. Rong Yu, Dr. Jing Tao and Dr. Jun Li for helpful discussions, Mengke Liu and Zongyuan Zhang for assisting with some experiments. 
The neutron scattering work at Rice is supported by the
U.S. DOE, BES under contract no. DE-SC0012311 (P.D.). 
The Ba(Fe$_{1-x}$Cu$_2$)$_2$As$_2$ single crystal synthesis work at Rice is supported by the Robert A. Welch Foundation Grants No. C-1839 (P.D.).
We acknowledge the support of the High Flux Isotope Reactor, a DOE Office of Science User Facility operated by ORNL 
in providing the neutron research facilities used in this work.

\end{document}